# SPICE, A Dataset of Drug-like Molecules and Peptides for Training Machine Learning Potentials

Peter Eastman[1], Pavan Kumar Behara[2], David L. Dotson[3], Raimondas Galvelis[4], John E. Herr[5], Josh T. Horton[6], Yuezhi Mao[1], John D. Chodera[7], Benjamin P. Pritchard[8], Yuanqing Wang[7,10], Gianni De Fabritiis[4,9], Thomas E. Markland[1]

[1]Department of Chemistry, Stanford University, Stanford, CA 94305, USA
[2]Department of Pharmaceutical Sciences, University of California, Irvine, CA 92697, USA
[3]The Open Force Field Initiative, Open Molecular Software Foundation, Davis, CA 95616, USA
[4]Acellera Labs, Doctor Trueta 183, 08005, Barcelona, Spain
[5]Department of Chemistry and Biochemistry, University of Notre Dame, Notre Dame, IN 46556, USA
[6]School of Natural and Environmental Sciences, Newcastle University, Newcastle upon Tyne NE1 7RU, United Kingdom
[7]Computational and Systems Biology Program, Sloan Kettering Institute, Memorial Sloan Kettering Cancer Center, New York, NY 10065, USA
[8]Molecular Sciences Software Institute, Virginia Polytechnic Institute and State University, Blacksburg, VA 24060, USA
[9]Computational Science Laboratory, Universitat Pompeu Fabra, Barcelona Biomedical Research Park (PRBB), Carrer Dr. Aiguader 88, 08003, Barcelona, Spain and ICREA, Passeig Lluis Companys 23, 08010 Barcelona, Spain.
[10]Graduate Program in Physiology, Biophysics, and Systems Biology, Weill Cornell Graduate School of Medical Sciences, New York, NY 10065, USA

## Abstract

Machine learning potentials are an important tool for molecular simulation, but their development is held back by a shortage of high quality datasets to train them on. We describe the SPICE dataset, a new quantum chemistry dataset for training potentials relevant to simulating drug-like small molecules interacting with proteins. It contains over 1.1 million conformations for a diverse set of small molecules, dimers, dipeptides, and solvated amino acids. It includes 15 elements, charged and uncharged molecules, and a wide range of covalent and non-covalent interactions. It provides both forces and energies calculated at the ωB97M-D3(BJ)/def2-TZVPPD level of theory, along with other useful quantities such as multipole moments and bond orders. We train a set of machine learning potentials on it and demonstrate that they can achieve chemical accuracy across a broad region of chemical space. It can serve as a valuable resource for the creation of transferable, ready to use potential functions for use in molecular simulations.

## Background & Summary

### Introduction

Machine learning potentials are an important, rapidly advancing tool for molecular simulation.[1] One creates a neural network or other machine learning model that takes atomic positions as input and produces potential energy and forces as output. The model is typically trained on a dataset of ground state potential energies, and possibly nuclear forces, produced with a conventional quantum chemistry method such as Density Functional Theory (DFT) or Coupled Cluster (CC). The model can be nearly as accurate as the original quantum chemistry method while being orders of magnitude faster to compute.[2,3] Machine learning potentials have been shown to be a practical tool for improving the accuracy of important calculations, such as protein-ligand binding free energies.[4,5] Many new architectures for machine learning potentials have been proposed in the last few years.[6–11]

Although the methodology for developing machine learning potentials is advancing quickly, much of the practical benefit has yet to be realized. There are very few general, pretrained, ready to use potentials available. A chemist or biologist wishing to perform calculations on a molecule cannot simply select a standard potential function, as they would a DFT functional or molecular mechanics force field. For machine learning potentials to fully realize their promise, the development of pretrained general (or transferrable) potential functions is essential. A user should be able to select a potential function and immediately start performing calculations with it.

A major factor currently limiting the creation of pretrained potentials is a lack of suitable training data. Any machine learning model is only as good as the data it is trained on. A person wishing to create a machine learning potential now has many architectures to choose from, but very few options for the data to train it on. Until a wider selection of high quality datasets becomes available, the field will be unable to take its place as a standard, widely used tool for molecular simulation.

In this article, we describe a new dataset called SPICE, which is short for Small-molecule/Protein Interaction Chemical Energies. As the name implies, this dataset is focused on capturing the energetics of molecular environments relevant to drug-like small molecules interacting with proteins. Potentials trained on this dataset should be able to produce accurate forces and energies for a wide range of molecules, and for the full range of conformations that occur in typical molecular dynamics simulations relevant to drug discovery. We hope it will become a valuable resource for the community, and that it will enable developers of new model architectures to create fully trained models that are ready to use.

The remainder of this article is organized as follows. The next section discusses the requirements a dataset should satisfy to be suitable for training machine learning potentials. We then review existing datasets and describe why none meet these requirements. We describe the design and construction of the SPICE dataset. We then present a proof of concept example to show that a machine learning potential function trained on the SPICE dataset can produce good accuracy. Finally we discuss plans for future updates to the dataset to further improve its accuracy and range of applicability.

**Dataset Requirements**

Our goal is to build accurate potential models useful for simulating drug-like small molecules interacting with proteins. Models trained on it should function as general purpose potentials, rather than specialized potentials constructed for a specific system: a user should be able to select a potential and immediately start using it on their molecule of interest. To achieve this goal, a dataset should meet the following requirements.

1. It must cover a wide range of chemical space.

The dataset should include data for a wide range of elements and chemical groups found in drug-like molecules and their protein targets. It should include charged and polar molecules. It should sample a variety of both covalent and non-covalent interactions.

2. It must cover a wide range of relevant conformational space.

Some existing datasets include only low energy conformations, or worse, only energy minimized ones. Molecular dynamics simulations and conformational search algorithms frequently explore higher energy conformations. A potential function must produce accurate results for all conformations encountered in these applications.

On the other hand, we explicitly do not include very high energy conformations encountered when forming or breaking covalent bonds. Our current goal is to simulate conformational changes rather than chemical reactions that involve making and breaking of bonds. A future version of the dataset may include additional data needed to accurately simulate chemical reactions.

3. It must include forces as well as energies.

Some existing datasets include only energies, not forces. The addition of forces enormously increases the information content of the dataset. For a molecule with N atoms, there are 3N force components and only one energy. Given the modest additional cost of computing forces, there is no reason not to include them. A dataset without forces is far less useful for training models than one that includes them.[12–14]

4. It should use the most accurate level of theory practical.

A machine learning potential can never be more accurate than the data it is trained on. Improving the accuracy of the training data immediately improves the accuracy of every model trained on it. One should therefore use the very most accurate level of theory practical, given the available computer resources and the amount of data to be generated.

5. It should include other useful information when possible.

Quantum chemistry calculations can very inexpensively produce other pieces of information along with the forces and energies: partial charges, multipole moments, etc. This information could potentially be useful in training various types of models, or in improving a potential model via multitask training approaches. We follow the principle that the dataset should include any information that is inexpensive to calculate and store, and that could potentially be useful.

6. It should be a dynamic, growing, versioned dataset.

Even after it is created, a dataset should continue to grow. Our plan is to continue generating data both to improve the accuracy of trained models and to expand their range of applicability. To allow reproducibility, we will regularly produce versioned releases of the dataset.

**Existing Datasets**

In this section we survey existing public datasets, and discuss why they are not sufficient for our requirements.

OrbNet Denali[15]

Of the existing datasets, this one comes closest to meeting our requirements. It includes 2.3 million conformations for a wide range of molecules. It includes 17 elements and charged as well as uncharged molecules. It includes both low and high energy conformations.

The most serious limitation of this dataset is that it includes only energies, not forces. This means that despite the large number of conformations, the total information content is quite limited, only a single number per conformation. Another issue is that, while it contains extensive sampling of covalent interactions, the sampling of non-covalent interactions is much more limited.

QMugs[16]

This is another large dataset containing nearly 2 million conformations for 665,000 chemically diverse molecules with up to 100 heavy atoms. It includes 10 elements and has polar molecules, but not charged ones. It includes only energy minimized conformations, which makes it unsuitable for training potential functions.

ANI-1[17], ANI-1x[18], and ANI-1ccx

ANI-1 is one of the largest public datasets with over 20 million conformations for 57,462 small molecules. Unfortunately, the coverage of chemical space is extremely limited. It contains only four elements and no charged molecules. This is insufficient for either drug molecules or proteins. It also includes only energies, not forces. The closely related ANI-1x and ANI-1ccx datasets do include forces for smaller numbers of conformations, but they otherwise share the same limitations.

QM7[19,20], QM8[21,22], and QM9[21,23]

These are also popular datasets for testing machine learning models. They have the same limitations as ANI-1: only five elements, no charges, and no forces. They have the further limitation of only including energy minimized conformations.

QM7-X[24]

This dataset improves on QM7 by sampling a variety of equilibrium and non-equilibrium conformations for each molecule. It also adds one more element, and reports a variety of physicochemical properties for each conformation. It nonetheless is insufficient for our needs, since it includes only six elements, only tiny molecules (up to seven heavy atoms), no charged molecules, and no sampling of non-covalent interactions.

DES370K and DES5M[25]

These datasets are based on a collection of approximately 400 very small molecules (up to 22 atoms) chosen to cover a wide range of chemical space. They are combined into approximately 370,000 dimer conformations computed at the CCSD(T) level of theory, and an additional 5 million dimer conformations computed at the SNS-MP2 level of theory. There are 20 elements and charged as well as uncharged molecules.

The energies in these datasets include only the intermolecular interaction energy. All intramolecular interactions are omitted. This makes them unsuitable for training machine learning potentials. As with the other datasets listed above, they contain only energies, not forces.

AIMNet-NSE[26]

A dataset of 295,000 conformations for molecules with between 13 and 16 heavy atoms was created for evaluating this model. It includes nine elements and contains both charged and neutral molecules. A much larger dataset of 6.44 million conformations for molecules with up to 12 heavy atoms was used to train the model, but it has not been made publicly available. Much of the data is for open-shell molecules, which is relevant to its original purpose of simulating chemical reactions but not to our purpose of simulating protein-ligand binding. It contains only energies, not forces.

## Methods

### Content of the SPICE Dataset

The SPICE dataset is made up of a collection of subsets, each designed to add particular information. For example, some focus on covalent interactions while others focus on noncovalent ones. Different subsets also contain different types of chemical motifs. The goal is that when all the subsets are combined into the complete dataset, it should have broad sampling of all types of interactions found in the systems we are interested in. The subsets are described below, along with high level descriptions of how they were created. For full detail on the data generation process, the scripts used to generate them are available online at https://github.com/openmm/spice-dataset.

Dipeptides

This subset is designed to provide broad coverage of covalent interactions found in proteins. It consists of all possible dipeptides formed by the 20 natural amino acids and their common protonation variants. That includes two forms of CYS (neutral or negatively charged), two forms of GLU (neutral or negatively charged), two forms of ASP (neutral or negatively charged), two forms of LYS (neutral or positively charged), and three forms of HIS (neutral forms with a hydrogen on either ND1 or NE2, and a positively charged form with hydrogens on both). This gives 26 amino acid variants, which can be combined to form 676 possible dipeptides. Each one is terminated with ACE and NME groups. A pair of CYS residues, each terminated with ACE and NME groups and connected to each other by a disulfide bond, is also included.

Dipeptides are sufficient to sample the full range of covalent interactions found in naturally occurring proteins (not counting post-translational modifications). Longer peptides allow a wider variety of non-covalent interactions, but have no additional covalent ones.

For each of the 677 molecules, the dataset includes 50 conformations of which half are low energy and half are high energy. To generate them, RDKit 2020.09.3[27] was first used to generate 10 diverse conformations. Each was used as a starting point for 100 ps of molecular dynamics at a temperature of 500K using OpenMM 7.6[28] and the Amber14[29] force field. A conformation was saved every 10 ps to produce 100 candidate high energy conformations. From these, a subset of 25 was selected that were maximally different from each other as measured by all atom RMSD.

Starting from each of the 25 high energy conformations, a nearby low energy conformation was created as follows. First, five iterations of L-BFGS energy minimization were performed. This was sufficient to bring it close to a local minimum but did not fully minimize it. Minimization was followed by 1 ps of molecular dynamics at 100K. This procedure yielded the 25 low energy conformations.

Solvated Amino Acids

This subset is designed to sample protein-water and water-water interactions. These are critical non-covalent interactions in protein simulations, so having good sampling of them is essential. It contains each of the 26 amino acid variants described above, terminated with ACE and NME groups and solvated with 20 TIP3P-FB water molecules.

The dataset includes 50 conformations for each one. To generate them, each amino acid was solvated with a 2.2 nm wide cube of water. 1 ns of molecular dynamics at 300K was run and a conformation was saved every 20 ps. For each conformation, the 20 water molecules closest to the amino acid were identified, as measured by the distance from the water oxygen to any solute atom, and the rest of the water was discarded.

PubChem Molecules

This subset includes a large and diverse set of small, drug-like molecules. We began by downloading from PubChem every substance record whose source was either BindingDB[30] (which contains small, drug-like molecules that bind to proteins) or ChemIDplus[31] (which contains molecules cited in National Library of Medicine databases). This is approximately 1.5 million records. Next they were filtered to only include records meeting all of the following criteria.

- The record contains only a single molecule.
- The molecule has between 3 and 50 atoms, including hydrogens.
- It does not include any elements other than Br, C, Cl, F, H, I, N, O, P, and S.
- There are no radical electrons.
- RDKit reports no sanitization errors while processing the molecule.
- It can be parametrized with the OpenFF 2.0.0 force field.[32]

Finally, we selected a subset of 14,644 molecules chosen to be maximally different from each other as measured by the Tanimoto similarity[33] between their ECFP4 fingerprints.[34] 50 conformations for each one were chosen using the same process described above for dipeptides.

DES370K Dimers

This subset includes all dimer structures from the DES370K dataset, except ones involving noble gas atoms which were omitted as out of scope. There are a total of 345,682 conformations that sample a very wide range of non-covalent interactions. By including these conformations, we gain extensive sampling of non-covalent interactions between a diverse set of chemical groups. They complement the dipeptides and PubChem molecules, which primarily provide information about covalent interactions.

DES370K Monomers

This subset includes each of the 378 monomers with more than one atom in the DES370K dataset. Each one has 50 conformations generated with the same process described above for dipeptides. It is included to supplement the DES370K dimers, because the latter largely consist of rigidly displacing monomers relative to each other without attempting to sample their internal motions.

Ion Pairs

This subset contains pairs of monatomic ions at varying distances from each other. DES370K already includes some data of this sort, but only at short distances and only for pairs with opposite charges. We therefore include all of the 28 possible pairs consisting of two ions chosen from Br$^-$, Cl$^-$, F$^-$, I$^-$, K$^+$, Li$^+$, and Na$^+$. Each pair is sampled at all distances between 2.5 and 7.5 Å, in increments of 0.1 Å.

See Table 1 and Figure 1 for a summary of the overall content of the dataset.

| Subset | Molecules | Conformations | Atoms | Elements |
|---|---|---|---|---|
| Dipeptides | 677 | 33850 | 26–60 | H, C, N, O, S |
| Solvated Amino Acids | 26 | 1300 | 79–96 | H, C, N, O, S |
| DES370K Dimers | 3490 | 345676 | 2–34 | H, Li, C, N, O, F, Na, Mg, P, S, Cl, K, Ca, Br, I |
| DES370K Monomers | 374 | 18700 | 3–22 | H, C, N, O, F, P, S, Cl, Br, I |
| PubChem | 14643 | 731856 | 3–50 | H, C, N, O, F, P, S, Cl, Br, I |
| Ion Pairs | 28 | 1426 | 2 | Li, F, Na, Cl, K, Br, I |
| Total | 19238 | 1132808 | 2–96 | H, Li, C, N, O, F, Na, Mg, P, S, Cl, K, Ca, Br, I |

**Table 1.** The overall content of the SPICE dataset. For each subset, the columns indicate 1) the number of molecules/clusters in the subset, 2) the total number of conformations, 3) the range of sizes spanned by the molecules/clusters in the subset, and 4) the list of elements that appear in the subset.

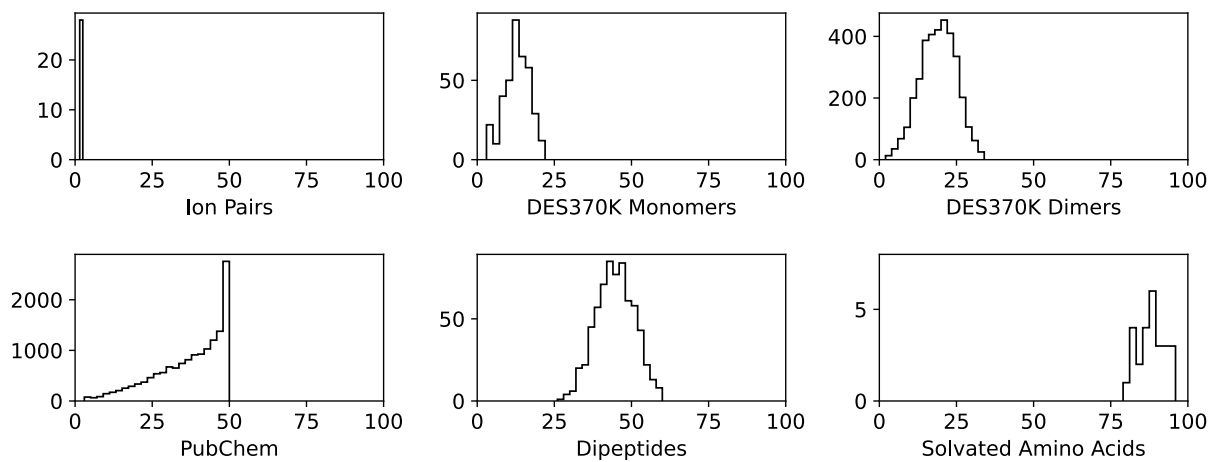

**Figure 1.** A histogram of the sizes of molecules in each subset. The horizontal axes are the numbers of atoms in a molecule, and the vertical axes are the numbers of molecules in each bin.

## Quantum Chemistry Calculations

Calculations were set up using OpenFF-QCSubmit,[35] and computation was managed with QCFractal server infrastructure.[36] The electronic structure code used for energy and gradient calculations is Psi4 1.4.1,[37] via the calculation backend QCEngine. Forces and energies for all conformations were calculated using DFT with the ωB97M-D3(BJ) functional[38,39] and def2-TZVPPD basis set.[40,41] This combination was chosen as the most accurate level of theory supported by Psi4 that was practical within our computational budget.[42–

[44] To get significantly higher accuracy, it would have been necessary to use either a double hybrid functional or a quadruple zeta basis set, either of which would be much more expensive.

In addition to forces and energies, the dataset includes a number of other properties for each conformation. Some of the more useful include: dipole and quadrupole moments; MBIS charges, dipoles, quadrupoles, and octopoles for each atom; Wiberg bond orders; and Mayer bond orders.

**Future Additions**

The dataset described in this article is the first release of SPICE. We intend that it will grow with time as we continue to add more data.

An important focus of future work will be to improve the accuracy of models trained on the dataset. This can be done with active learning.[45] One independently trains several models. To the extent that the training data constrains the model output, all the models should produce similar predictions. Any disagreement between models therefore serves as a measure of how well the training data constrains the output for a given conformation. One searches for molecules and conformations on which the models have the most disagreement. These are the ones that should produce the most benefit when they are added to the dataset.

Another goal is to extend the range of chemical space covered by the dataset. The initial release includes all of the most common elements found in drug molecules, but it is not exhaustive. Boron and silicon are top candidates for the next elements to add due to their appearance in drug molecules. We also could add training data for a wider range of biological molecules: nucleic acids, lipids, carbohydrates, proteins with post-translational modifications, etc., along with more extensive coverage of protonation and tautomeric states.

Another possible extension is to add data for high energy conformations in which covalent bonds are being formed or broken. This would enable the creation of reactive potential functions that can simulate the process of forming and breaking bonds.

## Data Records

The SPICE dataset has been deposited in Zenodo under accession number doi:10.5281/zenodo.7338495.[46] The data is contained in a single HDF5 file called SPICE-1.1.2.hdf5. It is organized as a single top level group for each molecule or cluster. Within each group are data fields for the different types of data available on the molecule. Detailed descriptions are available at the Zenodo record cited above.

## Technical Validation

**Training a Potential Energy Model**

In this section we train a set of machine learning potentials on the SPICE dataset. This is intended only as a proof of concept to illustrate one of the ways in which this dataset can be used. In a future publication we plan to explore this subject in far more detail, testing a variety of model architectures and developing one or more pretrained, ready to use potential functions. For the moment, our goal is only to demonstrate the value of the SPICE dataset by showing that it is possible to train high quality potential functions on it that cover a wide range of chemical space.

We train an ensemble of five models. All models use identical hyperparameters, varying only in the random number seed. We use the Equivariant Transformer model as implemented in TorchMD-NET as our model architecture.[8] The model includes 6 layers, each with 8 attention heads and a total embedding width of 128, a cutoff distance of 1 nm, and 64 radial basis functions. Because this is only a proof of concept, minimal hyperparameter tuning was done before selecting these parameters. We do not claim they are optimal; extensive hyperparameter tuning will be an important part of future work.

The best way of representing charged and polar molecules in machine learning models is still an open question. Several possible approaches have been proposed,[7,47,48] while other models avoid the problem simply by not supporting charges.[49] For this model we use a particularly simple mechanism. Rather than equating atom types with elements, as done in many other models, we take each unique combination of element and formal charge to be a different atom type. This leads to 28 atom types in total that are present in the SPICE dataset. They are listed in Table 2. In the equivariant transformer model, each atom type corresponds to a different embedding vector that can appear as an input to the first layer.

| Element | Charge | Instances | Element | Charge | Instances |
|---|---|---|---|---|---|
| H | 0 | 1594 | Mg | 2 | 1488 |
| Li | 1 | 3531 | P | 0 | 41528 |
| C | -1 | 5899 | P | 1 | 750 |
| C | 0 | 12545137 | S | -1 | 3350 |
| C | 1 | 1800 | S | 0 | 512526 |
| N | -1 | 11642 | S | 1 | 3945 |
| N | 0 | 2231039 | Cl | -1 | 7622 |
| N | 1 | 114621 | Cl | 0 | 246165 |
| O | -1 | 81548 | K | 1 | 6704 |
| O | 0 | 2235856 | Ca | 2 | 1587 |
| O | 1 | 1500 | Br | -1 | 4276 |
| F | -1 | 4033 | Br | 0 | 87927 |
| F | 0 | 376898 | I | -1 | 4344 |
| Na | 1 | 6536 | I | 0 | 21908 |

**Table 2.** The 28 atom types and the number of times each one appears in the complete SPICE dataset. The most common atom type, a neutral carbon, appears over 12 million times. The least common type, a positively charged phosphorus, appears only 750 times.

A limitation of this representation is that formal charges depend on covalent structure, and therefore atom types remain valid only so long as bonding patterns do not change. Because the initial version of SPICE is not designed to sample bond formation, this is not a problem for the current model. In addition, the assignment of formal charges is not always unique. That is not necessarily a problem, but it could potentially impact the accuracy of predictions in cases where the charges are ambiguous.

*Removal of atomic internal reference energies*

Approximately 99.9% of the energy of each conformation comes from the internal energies of the individual atoms. In principle the model should be able to learn this constant energy and separate it from the part that varies with conformation, but in practice it makes training far more difficult. It also is not relevant for molecular simulations. We therefore subtract the energies of the individual isolated atoms, training the model to reproduce formation energies rather than total energies.

*Filtering of strained molecules*

In training the models, we discovered that a small fraction of conformations are highly strained, and that they produce large errors that prevent the optimizer from learning. We therefore discard samples for which any component of the force on any atom is larger than 1 hartree/bohr. This constitutes approximately 2% of all samples. After removing them, we are left with 1,109,212 total samples. We randomly select 5% of the remaining samples (55,461) for use as a test set for each model, and use the rest as a training set. Each model uses a different random test set.

*Optimization of model parameters*

Each model was trained for 24 hours on four NVIDIA A100 GPUs, in which time it completed 118 epochs. Training uses an AdamW optimizer with batch size 128. The loss function is a weighted sum of the L2 loss for energies and forces, with a weight of 1 (kJ/mol)$^{-2}$ for the energy loss and 1 (kJ/mol/Å)$^{-2}$ for the force loss. The learning rate is initially set to 0.0005, then divided by 2 after any epoch on which the training loss fails to decrease.

The evolution of the loss function and learning rate during training are shown for all five models in Figure 2. In all cases, by the end of training the learning rate has decreased to a much lower value and the training loss has largely plateaued. This suggests the model has converged, and further training would be unlikely to produce significant improvement.

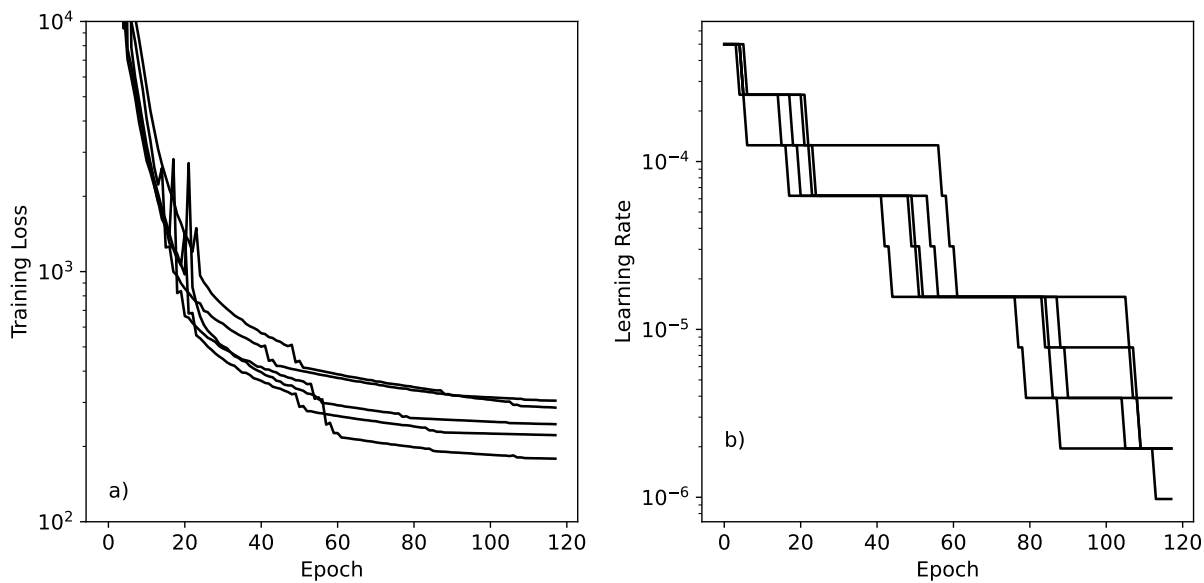

**Figure 2.** Evolution of a) the loss function on the training set and b) the learning rate during training for each of the five models. The loss function is a weighted sum of the L2 loss for energies and forces, with a weight of 1 (kJ/mol)$^{-2}$ for the energy loss and 1 (kJ/mol/Å)$^{-2}$ for the force loss.

The final value of the loss function on the training set and test set is shown in Figure 3 for each of the five models. In all cases, the difference between training loss and test loss is quite small, much less than the difference between models. Over the five models, the training loss ranges from 179 to 305 with a mean of 247. The test loss ranges from 195 to 318 with a mean of 264. This means the RMS error in predictions on test set samples is higher by only a factor of sqrt(264/247) = 1.03. This indicates that overfitting is not a problem, and the model should be nearly as accurate on novel conformations as on the ones it was trained on. The differences between models simply indicate that the optimizer was more successful at optimizing some models than others, and better training set accuracy directly translates to better test set accuracy.

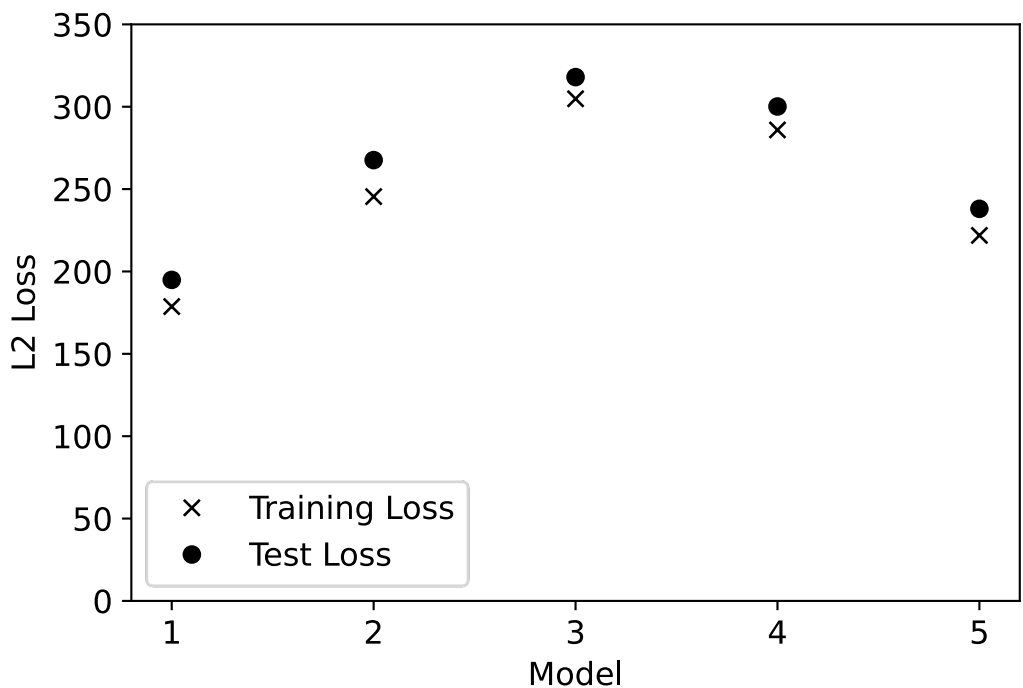

**Figure 3.** The final value of the loss function on the training set and test set for each of the five models. Each model uses a different random split for the training and test sets. The loss is a weighted sum of the L2 loss for energy and force predictions.

*Analysis of the trained models*

Having trained a set of models, we now analyze them to see what we can learn from them about the SPICE dataset. We emphasize once again that these models are only a proof of concept and not intended for production use. Our hope is that by understanding where they succeed and fail, we will learn about the likely behavior of other models trained on the same dataset, as well as gaining insight into the best ways to improve the models and expand the dataset to produce better accuracy.

We use the best of the five models (model 1 in Figure 3, with training loss 179 and test loss 195) to predict the energy of every conformation in the complete dataset (both training and test sets). Across the entire dataset, the mean absolute error (MAE) is 4.663 kJ/mol. The median absolute error is significantly lower, 2.912 kJ/mol, indicating that most predictions are highly accurate but a small fraction of samples with large errors increase the mean.

One possibility we consider is that our simple method of representing charges through atom types may not be sufficiently accurate, leading to low accuracy on charged or polar molecules. Figure 4 tests this hypothesis by plotting a) the MAE versus the total charge of each molecule, and b) the MAE versus the number of atoms with non-zero formal charge. There does appear to be some dependency. The error is lowest for molecules with 0 or 1 charged atoms, or with a total charge between -1 and +2. This suggests that more sophisticated ways of handling charges could potentially improve accuracy. On the other hand, the dependency is not monotonic, and many molecules with large numbers of charged atoms still have low error. We conclude that the charge representation is generally effective, although there certainly is room for improvement.

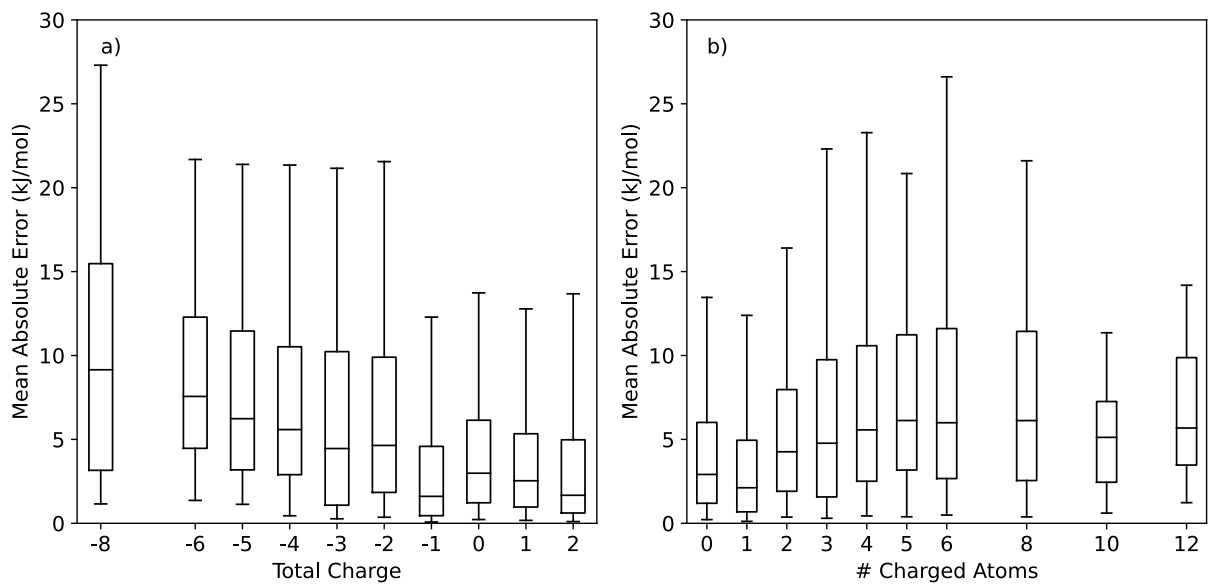

**Figure 4.** Absolute error as a function of a) the total charge of a molecule, and b) the number of atoms with non-zero formal charge in a molecule. The horizontal line indicates the median, the box spans the central 50% of samples, and the whiskers span 90% of samples.

Another possibility is that accuracy may depend on the size of the molecule. This is shown in Figure 5. Because the model predicts a separate energy for each atom, we expect the total error should generally increase with molecule size, and indeed this is the case. Across the full range, the absolute error seems to be roughly linear in the number of atoms. The analysis is complicated by the fact that samples of different size are qualitatively different from each other (see Table 1 and Figure 1). The smallest ones largely consist of DES370K dimers, intermediate size ones are largely PubChem molecules, and the largest ones are all solvated amino acids. Within each group, the dependence on size appears to be sublinear. A more thorough analysis of how accuracy varies with size is beyond the scope of this proof of concept and will be covered in a future publication.

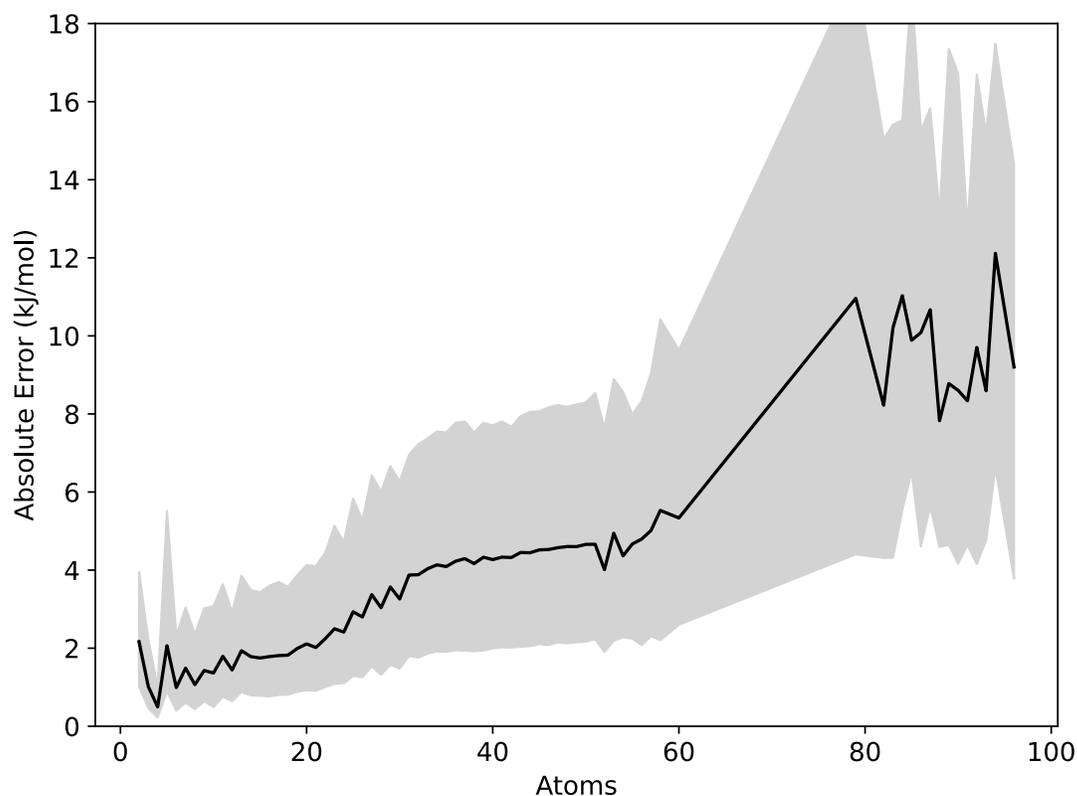

**Figure 5.** Absolute error as a function of the total number of atoms in a molecule. The line indicates the median for all molecules of a certain size, and the gray region contains the central 50% of samples.

*Molecules with large errors*

Another useful analysis is to compute the MAE across all conformations for each molecule and identify the molecules on which the model is least accurate. They are shown in Table 2 and Figure 6. They tend to contain unusual chemistries that are not typical of most molecules in the dataset. Of the 16 molecules with the largest errors, seven are phosphinic acids, while four others feature a carbon that forms two double bonds. This suggests a major source of error may be the limited data available for unusual chemistries. If so, it can be improved by identifying them and generating additional training data for those cases.

| SMILES | MAE (kJ/mol) |
|---|---:|
| O=C(O)CC(=O)C[PH](O)(O)CCc1c(Cl)cc(Cl)cc1OCc1ccccc1 | 608.8 |
| C/N=C/[PH](C)(O)O | 545 |
| CSC(C[PH](O)(O)C(C)=N)C(=O)O | 468.7 |
| CC(C[PH](O)(O)C1=CCCN1C(=O)OCc1ccccc1)C(=O)O | 437.5 |
| CCOC(OCC)[PH](O)(O)CC(=O)CN | 428.4 |
| NCCC[PH](O)(O)C=C1CC1 | 387.3 |
| O=C[C@H](NCC(P(=O)(O)O)[PH](O)(O)O)c1ccccc1 | 242.7 |
| CCCCCC(C)/N=C(/P(=O)(O)O)[PH](O)(O)O | 220 |
| CC(C)CC(=N)[PH](O)(O)C(C)N | 151.6 |
| C1=CCCC=C=CCCC=1 | 99.1 |
| Cn1c(=O)c2c(=O)c3[nH]cc[nH]c3c(=O)c2c1=O | 88.8 |
| Cc1[nH]c2c(=O)c3c(c(=O)c=2[nH]c1C)C1C=CC3CC1 | 83.5 |
| c1cc2c3c(ccc4c3c1=NSN=4)=NSN=2 | 78.9 |
| C=C=C1CCCCC1 | 77.7 |
| CC=C=C(C)CN | 77.2 |
| C=C=Cn1ccc2c(Oc3ccc(N)cc3)ncnc21 | 77.2 |

**Table 2.** The 16 molecules in the dataset with largest mean absolute error across all conformations.

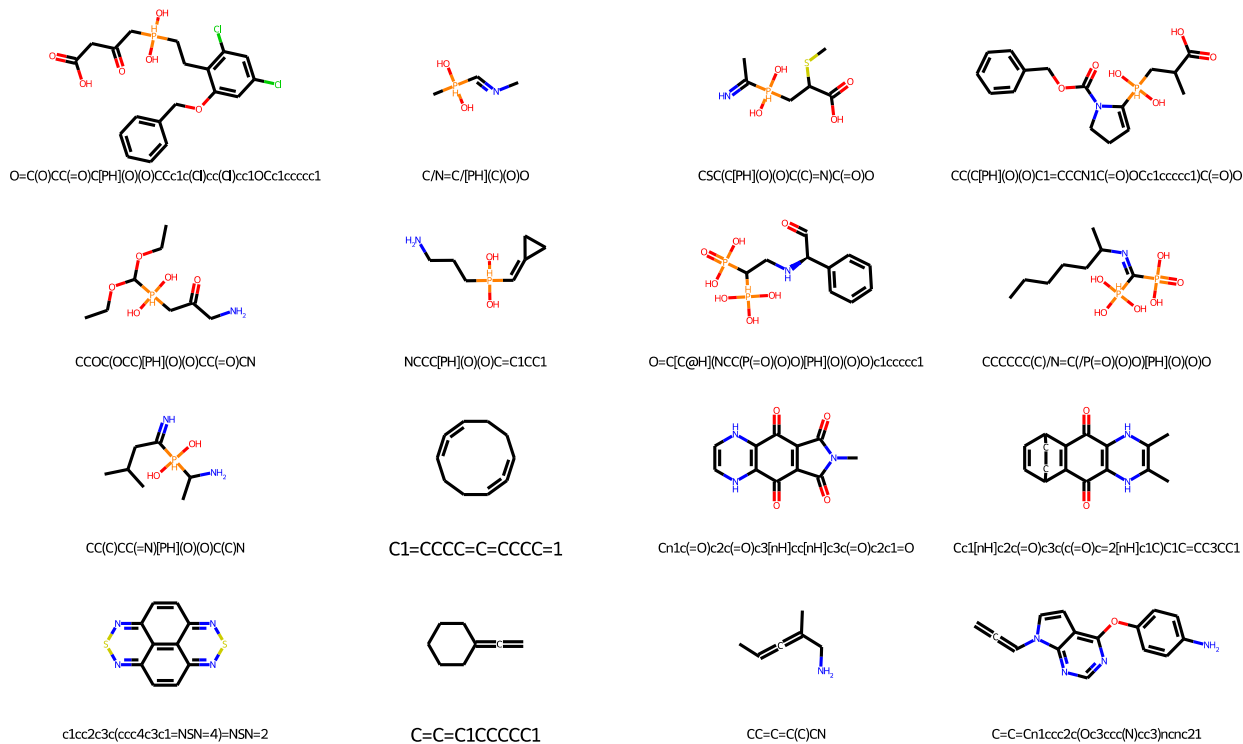

**Figure 6.** The 16 molecules in the dataset with largest mean absolute error across all conformations.

Despite the limitations, these results are excellent for a proof of concept. The majority of conformations already have errors below the 1 kcal/mol (4.184 kJ/mol) level commonly used as a standard for chemical accuracy.[50] With further hyperparameter tuning, more sophisticated handling of charges, and additional data for less common chemical motifs, we anticipate they can be significantly improved. That will be the subject of a future paper.

## Code Availability

The scripts used in the generation of the SPICE dataset are available online at https://github.com/openmm/spice-dataset. The trained equivariant transformer models and the hyperparameters used to train them are available at https://github.com/openmm/spice-models/tree/main/five-et.

## Author contributions

Peter Eastman: Conceptualization, Investigation, Software, Formal Analysis, Data Curation, Writing – Original Draft
Pavan Kumar Behara: Investigation, Data Curation, Writing – Review & Editing
David L. Dotson: Investigation, Data Curation
Raimondas Galvelis: Investigation, Writing – Review & Editing
John E. Herr: Methodology
Josh T. Horton: Investigation
Yuezhi Mao: Methodology, Writing – Review & Editing
John D. Chodera: Conceptualization, Resources, Funding Acquisition, Writing – Review & Editing

Benjamin P. Pritchard: Investigation
Yuanqing Wang: Investigation
Gianni De Fabritiis: Conceptualization, Writing – Review & Editing
Thomas E. Markland: Conceptualization, Resources, Funding Acquisition, Supervision, Writing – Review & Editing

## Acknowledgements

Research reported in this publication was supported by the National Institute of General Medical Sciences of the National Institutes of Health under award number R01GM140090 (JDC, TEM, PE, GdF) and R01GM132386 (JDC, PKB, YW). BPP acknowledges support from the National Science Foundation under award number CHE-2136142.

## Competing interests

JDC is a current member of the Scientific Advisory Board of OpenEye Scientific Software, Redesign Science, Ventus Therapeutics, and Interline Therapeutics, and has equity interests in Redesign Science and Interline Therapeutics. The Chodera laboratory receives or has received funding from multiple sources, including the National Institutes of Health, the National Science Foundation, the Parker Institute for Cancer Immunotherapy, Relay Therapeutics, Entasis Therapeutics, Silicon Therapeutics, EMD Serono (Merck KGaA), AstraZeneca, Vir Biotechnology, Bayer, XtalPi, Interline Therapeutics, the Molecular Sciences Software Institute, the Starr Cancer Consortium, the Open Force Field Consortium, Cycle for Survival, a Louis V. Gerstner Young Investigator Award, and the Sloan Kettering Institute. A complete funding history for the Chodera lab can be found at http://choderalab.org/funding.